\documentclass{elsarticle}
\usepackage{epsfig}
\usepackage{a4wide}

\def\lb#1{\if 1#1 \ln\beta \else \ln^#1\beta \fi}
\def\lt#1{\if 1#1 \ln 2 \else \ln^#1 2 \fi}

\newcommand{\be}{\begin{equation}}
\newcommand{\ee}{\end{equation}}
\newcommand{\ba}{\begin{eqnarray}}
\newcommand{\ea}{\end{eqnarray}}

\newcommand{\ep}{\epsilon}

\begin{document}


 \vspace{\baselineskip}

\title{
Next-to-leading order QCD effects and the top quark mass measurements at the LHC
}

    \author{Sandip Biswas, Kirill Melnikov and Markus Schulze}

    \address{
Department of Physics and Astronomy,
Johns Hopkins University,
Baltimore, MD, USA}


    \begin{abstract}
      \noindent
It is anticipated that a number of  techniques to measure
the top quark mass  at the LHC will yield  $m_{\rm top}$ with
uncertainties  of about $0.5-1$ percent. These uncertainties
are  mostly theoretical; they  are usually estimated
using  parton shower Monte Carlo programs whose reliability at this
level of precision  is difficult  to assess.
The goal of this paper is to contrast those estimates 
with  the 
results  of NLO QCD computations 
for  a few observables, often  
discussed in the context of  high-precision top quark mass measurements 
at the LHC.  In particular, we study the NLO QCD 
corrections to the invariant mass distribution of a charged lepton and
a $B$-meson in lepton+jets channels. In the dilepton channel
we  investigate the invariant mass distribution of a charged lepton 
and a $b$-jet, the average energy of the two
leptons and  the average energy of the $b$-jets from top decays.

      \end{abstract}

    \maketitle

\section{Introduction}

Measurements of  the top quark mass with highest  possible precision 
are useful for constraining physics beyond the Standard Model through
precision electroweak tests. In spite of spectacular Tevatron 
results \cite{TevatronCDF,TevatronD0}, 
the motivation to continue top quark measurements will 
remain strong even at the era of the 
LHC since one would like to see a consistency between
direct and indirect evidences for New Physics. Absence of
such a consistency will  be  a strong indication that our understanding
of emergent New Physics is incomplete.

 ATLAS and CMS plan to employ a  variety of methods
for measuring  the top quark mass
\cite{tdrs,Beneke:2000hk,rev1, rev2, rev3,Bernreuther:2008ju}.
These methods can be divided into
two classes. The first class includes so-called matrix element methods.
The idea is to  fit the top quark mass by adjusting
its value, to best describe
various kinematic features of $t \bar t$ events, using squared leading
order matrix elements as probability density functions.
Such  methods typically
lead to very small uncertainties in the top quark mass because
nearly all the information about the events is utilized. However,
the drawback of these methods is that  it is  hard
to estimate the theoretical uncertainty in the top quark mass
obtained in that way.

The second  class includes determination of  the top quark mass from
kinematic distributions that are sensitive to the value of $m_t$.
It is interesting that, up to now,
all analyses of  such distributions including their sensitivity to the
top quark mass and the theoretical uncertainty in $m_t$  have been
performed \cite{tdrs,Beneke:2000hk,rev1, rev2, rev3,Bernreuther:2008ju}
using parton shower event generators,
such as HERWIG \cite{hw} and PYTHIA\footnote{Even determination of the top quark
mass from the total cross-section for top quark pair production requires computation
of the acceptance since the total cross-section is never measured.
Such acceptances are routinely computed using parton shower event generators.}
\cite{pt}.   
As the result
of such studies,  it is often claimed that high precision in
the top quark mass measurement can be achieved.   Unfortunately,
it is not clear how reliable   such conclusions are since,
by construction,
parton showers  {\it can not guarantee} high  precision for
a generic observable due to their approximate nature. It is therefore
important to find alternative ways to estimate theoretical
uncertainties
in the description of  relevant kinematic distributions,
since those uncertainties impact directly
the precision of $m_t$  which can be expected from those measurements.

In this regard, we  point out that some kinematic
distributions that are expected to  be used for the top quark mass 
measurements,
can be computed in perturbative QCD. Typically, those distributions
involve top quark decay products; their  computation at leading
order in perturbative QCD is  straightforward.
On the other hand,   precision requirements on $m_t$
make it necessary to go to higher orders in the perturbative expansion
and, in spite of the fact that NLO QCD corrections to $t \bar t$ pair
production have been known for about  twenty years
\cite{nde,bnmss}, only recently NLO QCD
corrections to  top quark pair production {\it and} decay with all the spin
correlations included
 became available \cite{ms,bernsi,mcfm}.
The availability of NLO QCD corrections
to realistic final states and observables is a necessary
pre-requisite for the high-precision
analyses of kinematic distributions relevant for the top quark mass
determination. Higher-order corrections are also 
important from a more theoretical viewpoint 
since they allow us to  distinguish between the mass parameters 
defined in different renormalization schemes. In the context of 
$t \bar t$ production at the Tevatron, this issue was recently discussed 
in Ref.~\cite{moch}. Throughout this paper, we employ the pole 
mass of the top quark, for convenience.

The goal of this paper is to use the  computation
reported in Ref.~\cite{ms} as a starting point
to study some observables relevant
for the top quark mass determination
at  leading and next-to-leading order in QCD perturbation  theory
and to investigate
their sensitivity to input parameters. We will study  four
observables in this paper.
In Section~\ref{sect5} we
discuss  the invariant mass
distribution of a lepton and a $B$-meson from the top quark decay; this
is a simplified version of an observable suggested in
Ref.~\cite{kharch} for the top quark mass measurement.  Analysis of
$m_{Bl}$ distribution with high precision requires computation of
NLO QCD radiative corrections  to the {\it exclusive} decay $t \to B + X+l^+
+\nu$ which we report in  Section~\ref{sect5}.
In Section~\ref{dilepton}
we investigate the
invariant mass distribution of the $b$-jet and the charged lepton,  and
the distributions of the sum of energies of the
two leptons and the sum of energies of the  two
$b$-jets in the dilepton channel.
Calculation of those distributions is performed
using results reported in Ref.~\cite{ms}. We conclude in Section~\ref{conc}.

\section{Measurement of $m_t$  in top
decays to the  final state with identified $B$-meson}
\label{sect5}

It was pointed out in Ref.~\cite{kharch} that the top quark mass
can be accurately measured   by studying  top quark decays to
an exclusive hadronic state. For example,
one may consider the process
$pp \to (t \to W^++b \to W^+ + J/\psi)  + (\bar t \to W^- + \bar b)$
and require that the $W^-$ decays hadronically, $W^+$ decays leptonically,  $J/\psi$ decays into
a pair of leptons and $ \bar b$ decays into a lepton (inside the jet)
as well.  Then,
one may use  the invariant mass distribution
of a $J/\psi$ and an isolated  lepton
to determine the top quark mass. The requirement of a
large number of leptonic decays  reduces the rate significantly.
However, it also
reduces the combinatorial  background from
the incorrect pairing of the $J/\psi$ and a lepton. In addition,
since no jet measurements are involved,  the measurement is 
insensitive to jet energy scale  uncertainties.
As the result,  a very accurate
reconstruction of  the invariant mass  $m_{J/\psi l}$ and the
measurement of the top quark mass become possible.
It is expected \cite{kharch,cms,cmes}
that ${ \cal  O }(1~{\rm GeV})$  
error on the top quark mass 
can be achieved in such  measurement. The only (serious)
drawback of this method
is that very large luminosity -- about $100~{\rm fb}^{-1}$ -- is required
since the rate is suppressed due to  all
(semi)leptonic branching  fractions involved.   However, it was pointed
out in Ref.~\cite{cmsnote} that the situation can be improved by
giving up the requirement of the leptonic decay of the
$\bar b$-quark. In this case, combinatorial background increases but remains
manageable, and  the luminosity needed to reach uncertainty
of about $1.5~{\rm GeV}$  in the top quark mass is reduced to about
$20~{\rm fb}^{-1}$ \cite{cmsnote}.

The small uncertainty in the top quark mass that can, potentially,
be achieved in those measurements is very attractive. It also
sets the bar  for other methods
of the top quark mass measurement, planned at the LHC.
It is therefore peculiar that the analyses
in Refs.~\cite{kharch,cms,cmes} are  performed by using parton shower
event generators to describe production of top quarks and their decays.
The uncertainty in the extracted value of $m_t$  is estimated
in those references
by comparing results obtained using different parton shower event
generators, such as PYTHIA and HERWIG \cite{cmes}
or even different  versions of HERWIG
\cite{cms}. It is  possible  that parton shower event generators
give reasonable description of the required mass distribution
and that the resulting error estimates of  the top quark mass
are trustworthy. However, it is important  to check this,
given the potential importance
of the top quark mass measurement. To this end,  it is
useful  to look for alternative  ways to describe
the process $pp \to (t \to W^++b \to W^+ + J/\psi)
+ (\bar t \to W^- + \bar b)$, to ensure that the current understanding
of top quark decays and $b \to B$ fragmentation is consistent
with a very small error on $m_t$ that is claimed to be 
 achievable through $m_{J/\psi l}$ measurement.

One way to achieve that is to avoid using parton showers and,
instead, to compute  the $m_{J/\psi l}$ invariant mass distribution
in the process
$pp \to (t \to W^++b \to J/\psi)
+ (\bar t \to W^- + \bar b)$
in fixed-order perturbative QCD.
We describe how this can be done
in this Section.  To simplify the problem,
we follow  Refs.~\cite{cms,cmes} in that   we do not include
the decay of a $B$-meson to a $J/\psi$ meson, but  only consider
a $b$-quark fragmentation into a $B$-meson.  This is a reasonable
first step because decays of $B$-mesons to $J/\psi$-mesons are
well-studied at $B$-factories.
The energy spectrum of $B$-mesons in top decays can be
computed using the $b \to B$ fragmentation function formalism
\cite{nasonmele} which allows systematic inclusion of higher-order
QCD effects. The observable that we study in this Section
is the invariant mass distribution
of the $B$-meson and the lepton from the associated $W$-decay.

The NLO QCD calculation of the
$B$-meson energy spectrum in top quark decays  
was performed in Ref.~\cite{cm} 
within the $b \to B$
fragmentation  function formalism.  However, the
results of that reference can not be used directly for our
purpose since  leptons from $W$-decays were integrated over. Because
the primary object of our study is the invariant mass of a lepton from the
$W$ decay and a $B$-meson from the $b$-fragmentation and since we would
like to be able to impose kinematic
constraints on top quark decay products,
we require a calculation of the  NLO QCD corrections
that  is exclusive inasmuch as the  top quark decay products are concerned.

To perform  such a calculation, we  employ the dipole subtraction formalism
of Ref.~\cite{catani}. We point out, however, that we have an identified
hadron in the final state.  Hence, care is required
when the dipole subtraction
formalism is applied. In principle,  Ref.~\cite{catani} does describe the
construction of the subtraction terms for  such a situation,
but  since we deal here with decay kinematics and since massive particles are
involved, we adopt a slightly different approach.
As our starting point,
we take subtraction terms constructed specifically for
top quark
decays in Ref.~\cite{ekt}.  We modify the subtraction procedure
slightly, to allow for the identified hadron  in the final state, and
obtain a fully differential description of the decay
$t \to { l^+ \nu} + B   + X$ through NLO QCD.

\subsection{Calculation of radiative corrections
to $t \to l^+ \nu + B + X$.}
In this Section, computation of the radiative corrections to the
decay   $t \to (W^+ \to l^+ \nu) + B + X$ is described. We assume that
the $B$-meson is produced  by the fragmentation of the massless $b$-quark. The
$W$ boson is on the mass shell.
 We denote by $x$
the fraction of energy carried away by the $B$-meson
in the top quark rest frame\footnote{Except for the $B$-meson,
we denote particles and their momenta by the same label. We hope that
this fact does not cause any confusion.}
\be
x = \frac{2t p_{B}}{m_t^2(1-r^2)},\;\;\;\; r^2= \frac{m_W^2}{m_t^2}.
\ee

The differential decay rate reads
\be
 \frac{{\rm d} \Gamma_{B}}{{\rm d} x}  =
\int \limits_{x}^{1}
\frac{{\rm d} \xi}{\xi}\; \frac{{\rm d} \Gamma_{b}}{{\rm d} \xi} \;
D\left ( \frac{x}{\xi} \right ),
\label{eq4_1}
\ee
where $\xi = 2tb/(m_t^2(1-r^2))$, ${\rm d}\Gamma_{b}/{\rm d} \xi$
is the differential decay rate for the partonic decay
$t \to W + b +X$ and $D(x)$ is the fragmentation function
for $b \to B$.  One can restore the dependence on other partonic
variables in Eq.~(\ref{eq4_1}) because collinear fragmentation does not
affect them. Therefore,  
Eq.~(\ref{eq4_1}) is a starting point for the computation
of various kinematic  distributions for top quark
decays  to final states with  jets and an identified $B$-meson.

Our goal is to compute these distributions
through  next-to-leading order in perturbative QCD.
To this end, the partonic decay width ${\rm d} \Gamma$
is expanded in series of the strong coupling constant
$\alpha_s$
\be
{\rm d} \Gamma_b = {\rm d} \Gamma_b^{(0)} + {\rm d} \Gamma_b^{(\mathrm{V})}
+ {\rm d} \Gamma_b^{(\mathrm{R})} + {\cal O}(\alpha_s^2),
\ee
where the three terms refer to leading order decay rate and
virtual and real contributions to the NLO decay rate, respectively.
Because   ${\rm d} \Gamma_b^{(0)}$ and  ${\rm d} \Gamma_b^{(\mathrm{V})}$
have two-body  final states\footnote{We count decay
products of a $W$ boson as a single particle.},
the $b$-quark in that decay has maximal energy.
This implies that ${\rm d}\Gamma_b^{(0,\mathrm{V})}$ are
proportional to a delta-function of $\xi$
\be
{\rm d} \Gamma_b^{(0)} \sim {\rm d} \Gamma_b^{(\mathrm{V})} \sim \delta (1-\xi),
\ee
and  we can write
\ba
 \Gamma_{B}   &=&
\int {\rm d} x 
 \left \{ {\rm d} \Gamma_b^{(0)} + {\rm d}\Gamma_b^{(\mathrm{V})} \right \}
D(x) F_{J,2}(\{p\})
\nonumber \\
&& 
+ \int {\rm d} x \; 
{\rm d} \Gamma_b^{(\mathrm{R})}\; \frac{m_t^2(1-r^2)}{2bt}\;
D \left ( \frac{x m_t^2(1-r^2)}{2bt} \right )
F_{J,3}(\{p\}).
\ea
Note that we introduced the ``measurement function'' $F_{J,n}$ to indicate
external constraints that are applied to a $n$-particle final state.
The measurement function depends on the momenta of final state
particles, including the momentum of the $B$ meson. It
is assumed to satisfy the usual requirements of infra-red and collinear safety.

Since virtual and real corrections are separately infra-red and collinear
divergent and since  the measurement function is arbitrary,
we need to set  up a calculation where all divergences
in ${\rm d}\Gamma^{(\mathrm{R})}$ and in ${\rm d}\Gamma^{(\mathrm{V})}$ are regulated
separately.  We construct  the necessary subtraction term below
following Ref.~\cite{ekt} closely.
We begin by considering  the matrix element that describes the real emission
process $t \to W+b + g$
\be
{\rm d} \Gamma_b^{(\mathrm{R})} \propto |{\cal M}_3|^2~{\rm d}\Phi^{(3)},
\ee
where ${\rm d}\Phi^{(3)}$ is the phase-space element for $W,b$ and $g$, 
and introduce  variables $z,y$ to parametrize the scalar products that
involve the gluon momentum
\be
bg = \frac{m_t^2}{2} (1-r)^2 y;\;\;\; tg = \frac{m_t^2}{2} (1-r^2) (1-z) .
\ee
The soft limit $g \to 0$ requires $z \to 1,\;\; y \to 0$. The collinear
limit $g \ne 0,\; gb \to 0$ corresponds to $y \to 0$.
We express the fraction of maximal energy (in the top quark rest frame)
carried by the $b$ quark in $t \to b +g +W$ decay
through $z$ and $y$ variables
\be
\frac{E_b}{E_{b,\rm max}}
= \frac{2tb}{m_t^2(1-r^2)} = f(z,y) =
z + \frac{(1-r)}{(1+r)}y.
\label{eq_ww}
\ee
Using Eq.~(\ref{eq_ww}), we find that $f(z,y) \to 1$ and
$f(z,y) \to z$ in the soft and collinear limits, respectively.

The  soft and collinear limits  motivate the
construction of subtraction counter-terms
for real emission corrections. Taking the difference  of the real
emission correction and the subtraction counter-term, we obtain
an integrable expression
\be
\frac{ |{\cal M}_3|^2}{f(z,y)} \; F_{J,3}(\{p \},p_{B})
D \left ( \frac{x}{f(z,y)}\right )
-
\frac{|\tilde { \cal  M}_3(\tilde p)|^2}{z}\;
F_{J,2}\left (\{ \tilde p\}, \tilde p_{B} \right )
D \left ( \frac{x}{z} \right ),
\label{eq671}
\ee
which explicitly involves the fragmentation function.
We emphasize that,
as  with any subtraction method, the counter-term is evaluated
for values of momenta
that differ from the momenta used in the evaluation of the
matrix element. In particular, the  $B$-meson
momenta are related to a particular
$b$-quark momenta in the following way
\be
p_{B} = \frac{x b}{f(z,y)},\;\;\;\;\;\;\;
\tilde p_{B} =  x \tilde b.
\label{eq1010}
\ee
In the soft limit $f(z,y) \to 1$, $z \to 1$ and $b \to \tilde b$,
whereas  in the collinear limit, $f(z,y) \to z$, $b \sim z t$ and
$\tilde b \sim t$.  It follows from Eq.~(\ref{eq1010}) that 
$p_{B}$ and ${\tilde p}_B$  coincide in both limits; of course, this is 
an important condition  for the proper work of the
subtraction counter-term.
For the subtraction matrix element $|\tilde {\cal M}_3|^2$, we employ
\cite{ekt}
\be
|\tilde { \cal  M}_3|^2 = |\tilde {\cal M}_2|^2 \, {\rm Dip}(z,y),
\ee
where
\be
{\rm Dip}(z,y) = C_\mathrm{F} g_s^2 \mu^{2\epsilon}
 \left ( \frac{1}{bg} \left ( \frac{2}{1-z} - 1 -z - \eta \epsilon (1-z)
\right ) - \frac{m_t^2}{(tg)^2}
\right ),
\ee
and $\tilde {\cal M}_2$ is the matrix element for $t \to b + W$.
The term proportional to $\eta$ distinguishes the
t'Hooft-Veltman ($\eta =1$)  and the four-dimensional
helicity  ($\eta = 0$)  regularization schemes. We need to employ
the t'Hooft-Veltman scheme
in our calculation since this is the scheme (combined with
the ${\overline {\rm MS}}$ subtraction) in which fragmentation functions
are extracted from the $e^+e^-$ data in Ref.~\cite{cm}.

The required
momentum mapping is constructed in Ref.~\cite{ekt}; we summarize
it here for completeness.  We need to map a three  particle final
state ($ t \to W+b+g$) onto two particle final state ($ \tilde t \to \tilde W
+ \tilde b$).
We require that the top momentum
does not change, $\tilde t = t$, so that
\be
t = \tilde W + \tilde b.
\ee
Since $W^2 = {\tilde W}^2 = m_W^2$, $W$ is a valid candidate
to be the four-momentum of the $W$ boson after mapping but
it has wrong  energy for the two-body
decay. To correct for that, we can make a Lorentz transformation
\be
\tilde W^{\mu} = \Lambda^{\mu}_{\nu} W^{\nu}.
\ee
The matrix $\Lambda^{\mu \nu}$ reads \cite{ekt}
\ba
\Lambda^{\mu \nu} && = g^{\mu \nu}
+ \frac{\sinh(x)}{\sqrt{(tW)^2-m_t^2 m_W^2}}\left (t^\mu W^\nu
- W^\mu t^\nu \right) \\
&& + \frac{\cosh(x) - 1}{(tW)^2 - m_W^2 m_t^2}
\left (tW(t^\mu W^\nu + W^\mu t^\nu) - m_W^2 t^\mu t^\nu - m_t^2 W^\mu W^\nu
\right ),
\ea
where
\be
\sinh(x) = \frac{1}{2m_t^2 m_W^2}
\left ( - (m_t^2 - m_W^2) tW + (m_t^2 + m_W^2)\sqrt{ (tW)^2 - m_W^2 m_t^2 }
\right ).
\ee
Applying the Lorentz transformation to $W$, we obtain a simple expression
\be
\tilde W = \alpha \left ( W - \frac{tW}{m_t^2}\; t \right )
+ \beta t,
\ee
where
\be
\alpha = \frac{\sqrt{(m_t^2-m_W^2)^2 - 4 m_W^2 m_t^2}}{2\sqrt{(tW)^2 -
m_W^2 m_t^2}},\;\;\;
\beta = \frac{(m_t^2 - m_W^2)}{2 m_t^2}.
\ee
Explicit knowledge of the matrix
$\Lambda^{\mu \nu}$ is required  to account for momenta
changes of the decay products of the $W$ boson.

We are now in position to discuss the integration of the subtraction term
over the unresolved phase-space. We use the phase-space factorization,
described in Ref.~\cite{ekt}, ${\rm d} \Phi^{(3)}
= {\rm d} \tilde \Phi^{(2)} {\rm d} \tilde g$, where
\be
\int {\rm d} \tilde g =
\frac{(1-r)^2 m_t^{2-2\ep}}{16\pi^2} \frac{(4\pi)^\ep}{\Gamma(1-\ep)}
\left (
\frac{1+r}{1-r}
\right)^{2\ep}
\int \limits_{0}^{1} {\rm d} z
\left ( r^2 + z(1-r^2) \right )^{-\ep} \int \limits_{0}^{y_{\rm max}} y^{-\ep}
(y_{\rm max} - y)^{-\ep}.
\label{eq567}
\ee
In Eq.~(\ref{eq567}), $\epsilon$ is the dimensional
regularization parameter and  $ y_{\rm max} = (1+r)^2 z(1-z)/(z+r^2(1-z))$.
We need to calculate
\be
I_{\rm dip} = \int {\rm d}{ \tilde g} \;
{\rm Dip}(z,y)\; z^{-1}\; D \left ( \frac{x}{z}  \right ).
\ee

The two integrals over $y$ that we need (through an appropriate
order in $\epsilon $) are
$$
\int \limits_{0}^{y_{\rm max}}
\frac{{\rm d} y}{y} y^{-\ep}(y_{\rm max} - y)^{-\ep}
= y_{\rm max}^{-2\ep} \left ( -\frac{1}{\ep} + \ep \frac{\pi^2}{6} \right),
\;\;\;\;\int \limits_{0}^{y_{\rm max}}
{\rm d} y  y^{-\ep} (y_{\rm max} - y)^{-\ep} =
y_{\rm max}^{1-2\ep} ( 1 + 2\ep).
$$

Upon integrating over $y$, we obtain
\ba
 I_{\rm dip} = &&
\frac{2g_s^2 \mu^{2\ep} C_\mathrm{F} m_t^{-2\ep}}{(4 \pi)^{2-2\ep} \Gamma(1-\ep)
(1-r^2)^{2\ep}} \int \limits_{0}^{1}
\frac{{\rm d} z}{z} D \left ( \frac{x}{z} \right ) (r^2 + z(1-r^2) )^{\ep}
 z^{-2\ep} (1-z)^{-2\ep} 
\nonumber \\
&& \times
\left \{ \left ( \frac{2}{1-z} - 1- z\right )
\left (-\frac{1}{\ep} + \ep \frac{\pi^2}{6} \right )
- \frac{2 z(1+2\ep)}{(z+r^2(1-z))(1-z)}
\right \}.
\label{fgz}
\ea

Because the $z$-dependent fragmentation function $D$ is present in
Eq.~(\ref{fgz}), we can not  integrate over $z$ analytically.
The best we can do is to extract infra-red  and collinear
divergences.
We find
\ba
I_{\rm dip} =
&& \frac{C_\mathrm{F} \alpha_s (1-r^2)^{-2\ep} }{2 \pi \Gamma(1-\ep)}
\left ( \frac{4\pi \mu^2}{m_t^2} \right )^\ep
\Bigg[ \frac{D(x) }{\ep^2}
- \frac{1}{\ep} \int \frac{{\rm d} z}{z}
D \left ( \frac{x}{z} \right )
\left ( \frac{2}{(1-z)_+} - (1+z) - \delta(1-z) \right )
 \nonumber \\
&& \left.
+D(x) \left (2- \frac{\pi^2}{6} \right )
+\int \frac{{\rm d} z}{ z} D \left ( \frac{x}{z} \right )
\Bigg\{ 4 \left [ \frac{\ln(1-z)}{1-z} \right ]_+
 - 2 (1+z) \ln (1-z)
\right.  \nonumber \\
&&
-  \left ( \frac{2}{(1-z)}
- (1+z) \right ) \ln \left( \frac{r^2+z(1-r^2)}{z^2} \right)
- \frac{2}{(1-z)_+} \frac{z}{r^2+z(1-r^2) }
\Bigg\}
\Bigg].
\label{eq_intdip}
\ea

The infra-red  and collinear divergences explicit
in the result for  the integrated dipole Eq.~(\ref{eq_intdip})
must cancel with the virtual corrections
{\it and} the ${\overline {\rm MS}}$ renormalization of
the fragmentation function. The sum of the leading
order decay rate and the one-loop virtual
correction reads
\be
{\rm d} \Gamma_{B}^{(0)} +
{\rm d} \Gamma_{B}^{(V)} \sim |\mathcal{M}_2|^2 \; I_{\rm virt} 
\; {\rm d} \Phi_2,
\ee
where
\be
I_{\rm virt}  =  D(x) \left [ 1
+ \frac{\alpha_s C_\mathrm{F} (1-r^2)^{-2\ep} }{2\pi \Gamma(1-\ep)}
\left ( \frac{4\pi \mu^2}{m_t^2} \right )^{\ep}
\left ( C_0 + \frac{C_1}{2} \frac{1-r^2}{1+2 r^2} \right )
\right ].
\ee
The functions $C_{0,1}$ read
\cite{ekt}
\ba
&& C_0 = -\frac{1}{\ep^2} - \frac{5}{2\ep}
- \frac{11+\eta}{2} - \frac{\pi^2}{6} - 2{\rm Li}_2(r^2)
- 2\ln (1-r^2)  - \frac{\ln (1-r^2)}{r^2}
\nonumber \\
&& C_1 = \frac{2}{r^2} \ln(1-r^2).
\ea

Taking the sum of $I_{\rm dip}$ and
$I_{\rm virt}$, we obtain
\ba
&& I_{\rm dip} + I_{\rm virt}
= D(x) + \frac{\alpha_s C_\mathrm{F}}{2\pi \Gamma(1-\ep)}
\left ( \frac{4\pi \mu^2}{m_t^2} \right )^{\ep} (1-r^2)^{-\ep}
\Bigg [ D(x) V(r)
\nonumber \\
&&
- \frac{1}{\ep} \int \frac{{\rm d} z}{z}
D \left ( \frac{x}{z} \right )  \tilde P_{qq}(z)
+\int \frac{{\rm d} z}{ z} D \left ( \frac{x}{z} \right )
\Bigg\{ 4 \left [ \frac{\ln(1-z)}{1-z} \right ]_+
 - 2(1+z) \ln (1-z)
\ \nonumber \\
&&
-  \left ( \frac{2}{(1-z)}
- (1+z) \right ) \ln \left( \frac{r^2+z(1-r^2)}{z^2} \right)
- \frac{2}{(1-z)_+} \frac{z}{r^2+z(1-r^2) }
\Bigg\}
\Bigg ],
\label{eq_dip_vv}
\ea
where $\tilde P_{qq} = 2/(1-z)_+ - (1+z) + 3/2\;\delta(1-z)$ and,
for $\eta = 1$,
\be
V(r) = -\frac{7}{2} - \frac{\pi^2}{3} - 2 {\rm Li_2}(r^2)
-\frac{5+4r^2}{1+2r^2}\ln(1-r^2).
\ee

We observe that Eq.~(\ref{eq_dip_vv}) contains collinear divergences.
To remove them, the fragmentation function $D(x)$ needs to be renormalized.
By convention, we use the $\overline {\rm MS}$ scheme. We obtain
\ba
&& \overline{I_{\rm virt}}+\overline{I_{\rm dip}}
= D(\mu, x) + \frac{\alpha_s(\mu) C_\mathrm{F}}{2\pi}
\left ( \;  D(\mu,x) V(r)
- \ln \left (\frac{\mu^2}{m_t^2(1-r^2)} \right )
 \int \frac{{\rm d} z}{z}
D \left (\mu, \frac{x}{z} \right )  \tilde P_{qq}(z)
\right. \nonumber \\
&& \left. 
+\int \frac{{\rm d} z}{ z} D \left ( \mu,\frac{x}{z} \right )
\left \{ 4 \left [ \frac{\ln(1-z)}{1-z} \right ]_+
 - 2(1+z) \ln (1-z)
\right. \right. \nonumber \\
&& \left. \left.
-  \left ( \frac{2}{(1-z)}
- (1+z) \right ) \ln \frac{r^2+z(1-r^2)}{z^2}
- \frac{2}{(1-z)_+} \frac{z}{r^2+z(1-r^2) }
\right \}
\right).
\label{eq_v_int}
\ea

Equation~(\ref{eq_v_int}) contains everything that is needed   to
compute  the contribution of the virtual corrections and the integrated
dipoles to the decay rate $t \to l^+ \nu + B   +X$.
These results should  be supplemented  with the
contribution of the real emission matrix elements, described by
Eq.~(\ref{eq671}). Combining Eq.~(\ref{eq_v_int}) and Eq.~(\ref{eq671}),
we can compute ${\cal O}(\alpha_s)$ correction to the
fully differential rate for $t \to l^+ \nu + B + X$. We then
interface the corrections to the decay, that we just described,
with the  production process, in the spirit of Ref.~\cite{ms}.  This
allows us to get a description of
$pp \to (t \to W^++b \to W^+ + J/\psi)
+ (\bar t \to W^- + \bar b)$
at leading and next-to-leading order, including the possibility to
apply kinematic  cuts to the final state particles.

\subsection{The fragmentation function}

The NLO QCD  calculation described in the previous Section
leads to  radiative corrections enhanced by the logarithm
of the ratio of the top quark mass and the factorization scale $\mu$.
We can choose $\mu \sim m_t$
to get rid of the logarithmically enhanced terms in the short-distance
partonic decay rate (cf. Eq.~(\ref{eq_v_int})).
However, by doing that, we face the challenge
of evaluating the fragmentation function $D(\mu,x)$ at a high value of
the factorization scale in spite of the fact that $b \to B$ fragmentation
is, intrinsically, the low-scale phenomenon.

The standard way to deal with the problem is to use the Altarelli-Parisi (AP)
equation
\be
\mu^2 \frac{\partial D_{b \to B}(x,\mu) }{\partial \mu^2}
= \sum_j
\int \frac{{\rm d} z}{z} P_{bj}\left ( \frac{x}{z},\alpha_s(\mu) \right )
D_{j \to B}(z,\mu),
\label{ap_eq}
\ee
to evolve the fragmentation function
to the required values of the factorization
scale $\mu \sim m_t$.
For the purpose of the NLO calculation, we
include ${\cal O}(\alpha_s)$ and ${\cal O}(\alpha_s^2)$ contributions
to the AP evolution kernel which leads to a resummation
of the leading and next-to-leading logarithms of the ratio of the factorization
scale and the $b$-quark mass.
Similar to what was done in the
previous studies, we neglect all off-diagonal contributions to the evolution
equation Eq.~(\ref{ap_eq}) and only keep there terms proportional
to $P_{bb}$ splitting function.

Solution of the AP equation requires an
initial condition, which is to say that $D_{b \to B}$ needs to be known
for some value of the factorization scale $\mu_0$.  Traditionally,
this is accomplished by fitting the fragmentation function at the 
scale
$\mu_0 \sim m_b$ to  data  on $e^+e^- \to b \bar b$
\cite{aleph,sld}.
Since  $\mu_0 \sim m_b$ is a  perturbative scale, 
we may attempt
to completely factorize perturbative and non-perturbative contributions,
by writing  the heavy quark fragmentation function as a convolution of the
perturbative fragmentation function $D_b(\mu,x)$ and the ``non-perturbative''
fragmentation function  $D_{\rm np}(x)$ \cite{nasonmele}
\be
D_{b \to B} (\mu,x) = \int \limits_{x}^{1} \frac{{\rm d} \xi}{\xi}
D_b(\mu,\xi) D_{\rm np}\left ( \frac{x}{\xi} \right ).
\ee
The perturbative fragmentation function receives contributions
from momenta comparable to the $b$-quark mass and
is therefore computable in perturbation theory.
At NLO QCD, the result reads \cite{nasonmele}
\be
D_b(\mu,x) = \delta(1-x) + \frac{\alpha_s(\mu)C_\mathrm{F} }{2\pi}
\left [\frac{1+x^2}{1-x} \ln\left( \frac{\mu^2}{m_b^2} \right)- 2 \log (1 - x) -1
\right ]_+ + {\cal O}(\alpha_s^2).
\label{eq:incond}
\ee
Note that the expansion parameter in Eq.~(\ref{eq:incond})
is  $\alpha_s \log \mu/m_b$. This observation
makes it clear that $D_b(\mu,x)$  has to be evaluated
at the scale $\mu \sim  m_b$.

In Ref.~\cite{cm} $D_{\rm np}(x)$ was determined from fits to the
$e^+e^- \to b \bar b$ data by ALEPH and SLD collaborations 
\cite{aleph,sld}. 
For numerical
calculations, we use results reported in that reference. We restrict
our attention to two types of non-perturbative
fragmentation functions
\be
D_{\rm np}  =
\left \{
\begin{array}{c}
x^{\alpha} (1-x)^{\beta}/B(\alpha+1,\beta+1),\;\;\; \\
(1+\delta)(2+\delta) (1-x) x^{\delta}.
\end{array}
\right.
\label{eq:ff_np}
\ee
It was shown in Ref.~\cite{cm} that the following values of
the parameters
\be
 \alpha = 0.66 \pm 0.13,\;\;
\beta = 12.39 \pm 1.04,\; \delta = 14.97 \pm 0.44,
\label{eq:ff_par}
\ee
lead to a good fit to the ALEPH data provided that {\it no soft gluon
resummation} is applied to the fragmentation function\footnote{
If soft gluon resummation in the perturbative fragmentation function
is employed, the preferred values of $\alpha, \beta$ and $\delta$ change,
see Ref.~\cite{cm} for details.}. We will use the
range of parameters shown above to estimate the sensitivity of
the extracted value of the top  quark mass to the  employed model
of the heavy quark fragmentation function.
We solve the AP evolution equation in a  standard way
by applying the Mellin transform since
the AP equation
becomes ordinary differential equation in the Mellin space.
The results that are required to perform the Mellin
transform  can be found in  Refs.~\cite{nasonmele,cm}.

\subsection{Results: $m_{Bl}$ distribution in top quark decays}

In this Section, we discuss our results for the $m_{Bl}$ spectrum,
as obtained within the perturbative fragmentation function
framework. We ignore all the
subtleties associated with the heavy quark production
mechanism and study the invariant mass of the lepton and the
$B$-meson as produced by the top quark decaying in isolation.
We quote results
at leading and next-to-leading order, but we need to clarify what
we mean  by that.  Indeed, a fragmentation function, is extracted
from data on  $e^+e^-$ annihilation to $B$-hadrons, using
a short-distance function for $e^+e^- \to b \bar b$,  computed through
a {\it particular}  order in perturbative QCD.
Therefore, if we change the short-distance
function by truncating it to leading order, we
get a different  fragmentation function.
This phenomenon is well-known from studies of parton
distribution functions that {\it do} change from one order in perturbation
theory to the other.  Unfortunately, information on how non-perturbative
fragmentation functions change when perturbative predictions
for $e^+e^- \to b \bar b  $ are truncated
at leading order are not available to us,  so that for our leading
order calculation we use the same  non-perturbative
fragmentation function $D_{\rm np}$, Eq.~(\ref{eq:ff_np}), as in next-to-leading
order computation. However,
for leading order computations, we neglect all the ${\cal O}(\alpha_s)$
corrections to partonic decay rate of the top quark and  the initial condition
$D_b$, Eq.~(\ref{eq:incond}),  and we
solve  the AP evolution equation and compute
the evolution of the strong coupling constant in the leading
logarithmic approximation.
To obtain numerical results reported below, we use
$\alpha_s(M_Z) = 0.130$ and $\alpha_s(M_Z) = 0.118$
for leading and next-to-leading computations, respectively.

In Tables~1~and~2, we show average values of
the invariant mass of the $B$-meson and the lepton
 $\langle m_{Bl} \rangle$ and the dispersion $\sigma_{m_{Bl}}$ of the
$m_{Bl}$ distribution at leading and
next-to-leading order in perturbative
QCD.   To arrive at those results, we calculate $\langle m_{Bl} \rangle $
for three values of the renormalization and factorization scales
$m_t,m_t/2,m_t/4$, changing them independently. We also use  two
different fragmentation functions, as explained in the previous Section.
For each parameter that one needs to describe the fragmentation
function, we do a calculation for its central value and for the central
value shifted by plus/minus the error quoted for that parameter.
As the result, we obtain $108$   values of $\langle m_{Bl} \rangle $
and $\sigma_{m_{Bl}}$
for each of the input values  of the top quark mass. We calculate
the mean and the error from these samples of $108$ numbers for
both $\langle m_{Bl} \rangle $ and $\sigma_{m_{Bl}}$  at leading
and next-to-leading order. In  Tables~1~and~2
those results are shown; the difference between the two Tables is
that an additional constraint  $m_{Bl} > 50~{\rm GeV}$ is
employed to obtain results in Table~2.

There are two immediate comments that one can make about
those results. First, we observe that
NLO QCD  corrections to $\langle m_{Bl} \rangle $
strongly depend on the applied  cut on the invariant mass.
For example, if no such cut is applied, the
shift from leading to next-to-leading order in $\langle m_{Bl} \rangle $ is about $2.5~{\rm GeV}$, whereas
if a $50~{\rm GeV}$ cut is applied, $\langle m_{Bl} \rangle $
shifts by $-0.3~{\rm GeV}$.
Second, the uncertainty in
$\langle m_{Bl} \rangle $ decreases by a factor between two and three,
when NLO QCD effects are included, indicating their importance for 
the high-precision top quark mass measurement.

It is interesting to compare the results of the computation reported
in this paper with the previous analysis where parton shower
event generators were employed \cite{cms,cmes}. We note that
results of those two references are not consistent; the reason is explained in
Ref.~\cite{cmes}.  We will therefore compare to the results in
Ref.~\cite{cmes}, where  $\langle m_{Bl} \rangle $ and higher
moments of $Bl$ invariant mass distribution
are  computed using HERWIG and PYTHIA.  The $B \to b$ fragmentation
functions were fitted in Ref.~\cite{cmes} to reproduce $B$-meson energy spectra
in $e^+e^-$ annihilation.  Systematic ${\cal O}(1~{\rm GeV})$
differences 
in values of $\langle m_{Bl} \rangle$ obtained with PYTHIA and HERWIG
were observed in \cite{cmes}, with PYTHIA results being lower.


\begin{table}[t]
\label{Jtab}
\vspace{0.1cm}
\begin{center}
\begin{tabular}{|c|c|c|c|c|}
\hline
$m_t$     &
$\langle m_{Bl}\rangle,\;{\rm LO} $ &
$\langle m_{Bl}\rangle,\;{\rm NLO}$ &
$\sigma_{m_{Bl}},\;{\rm LO}$  &
$\sigma_{m_{Bl}},\;{\rm NLO}$ \\ \hline \hline
$171$ & $73.51 \pm 1.87$&
$76.03 \pm 0.61$&
$31.46 \pm 0.12$&
$29.21 \pm 0.29$\\ \hline
$173$ & $74.71 \pm 1.90$&
$77.24 \pm 0.62$&
$31.92 \pm 0.12$ &
$29.63 \pm 0.29$\\ \hline
$175$ &
$75.91 \pm 1.93$ &
$78.44 \pm 0.63$ &
$32.37 \pm 0.13$ &
$30.04\pm 0.30$ \\ \hline
$177$ &
$77.10 \pm 1.95$  &
$79.64 \pm 0.63$  &
$32.82 \pm 0.13$  &
$30.46\pm 0.30$ \\ \hline
$179$ &
$78.29 \pm 1.98$ &
$80.84 \pm 0.64$ &
$33.26 \pm 0.13$ &
$30.87\pm 0.30$
 \\ \hline
\hline
\end{tabular}
\caption{The estimate of the
average value of the $B$-meson-lepton invariant
mass and its  dispersion at leading and next-to-leading order, in
dependence of the top quark mass. The top quark mass and all the results
are in GeV. Decay of an isolated top quark is considered.}
\vspace{-0.1cm}
\end{center}
\label{table:n1}
\end{table}

\begin{table}[t!]
\label{Jtabb}
\vspace{0.1cm}
\begin{center}
\begin{tabular}{|c|c|c|c|c|}

\hline
$m_t$ &
$\langle m_{Bl}\rangle,\;{\rm LO}$ &
$\langle m_{Bl}\rangle,\;{\rm NLO}$ &
$\sigma_{m_{Bl}},\;{\rm LO}$ &
$\sigma_{m_{Bl}},\;{\rm NLO}$ \\ \hline \hline
$171$ &
$87.51 \pm 1.04$&
$87.20\pm 0.43$&
$22.17 \pm 0.23$ &
$21.28\pm 0.17$ \\ \hline
$173$  &
$88.53 \pm 1.07$ &
$88.22\pm 0.43$ &
$22.68 \pm 0.24$ &
$21.77\pm 0.17$ \\ \hline
$175$ &
$89.56 \pm 1.10$ &
$89.25 \pm 0.44$ &
$23.19\pm 0.24$ &
$22.25\pm 0.18$ \\ \hline
$177$ &
$90.58 \pm 1.13$ &
$90.29 \pm 0.45$ &
$23.69 \pm 0.24$ &
$22.73\pm 0.18$ \\ \hline
$179$ &
$91.61 \pm 1.15$ &
$91.32 \pm 0.46$ &
$24.20 \pm 0.24$ &
$23.22\pm 0.18$
 \\ \hline
\hline
\end{tabular}
\caption{The average value of the
invariant mass $\langle m_{Bl} \rangle$ and its dispersion,
evaluated with  the cut
on the invariant mass   $m_{Bl}>50~{\rm GeV}$. 
The top quark masses and all the results are in GeV.
Decay of an isolated
top quark is considered.}
\vspace{-0.1cm}
\end{center}
\label{table:n2}
\end{table}

We find that the NLO QCD result for $\langle m_{Bl} \rangle $
and $\sigma_{m_{Bl}}$ are close to the  results obtained
with parton showers.  Nevertheless, the difference is
not negligible, given the expected precision of the top quark
mass measurement. By comparing our results with that of
Ref.~\cite{cmes}, we find that the average values of
$\langle m_{Bl} \rangle $ computed through NLO QCD is about
$2.4~{\rm GeV}$ lower  than $\langle m_{Bl} \rangle $
obtained with HERWIG and only $1.1~{\rm GeV}$
lower than $\langle m_{Bl} \rangle $ obtained PYTHIA.
On the other hand, the dispersion $\sigma_{Bl}$
that we compute through  NLO QCD, differs
by $1~{\rm GeV}$ from PYTHIA  
and by $0.5~{\rm GeV}$ from HERWIG results.

\begin{figure}[t!]
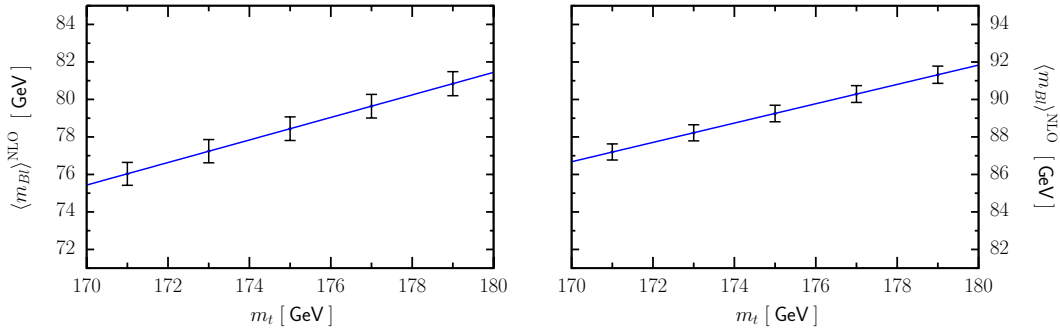
 
\begin{center}
\scalebox{0.45}{\input{MLB_mt_1.tex}} \hspace{5mm}
\scalebox{0.45}{\input{MLB_mt_2.tex}}
\vspace{3mm}
\caption{Results of the linear fit to $\langle m_{Bl} \rangle^{\rm NLO}$
are shown. Left panel -- no cut on $m_{Bl}$ is applied. Right panel --
$m_{Bl} > 50~{\rm GeV}$ cut is applied. In both cases, decays
of isolated top quarks are considered.
}
\label{fig1}
\end{center}
\end{figure}

The results for $\langle m_{Bl} \rangle$
displayed  in Tables~1~and~2 can be described by a linear function
of the top quark mass. We present the results of such
a fit for the two cases -- with and without
a cut on $m_{Bl}$ in Fig.~\ref{fig1}. We find
\ba
&& \langle m_{Bl} \rangle ^{\rm NLO} = 0.601 m_t - 26.7~{\rm GeV},
\;\;\;\delta_{\rm rms} = 0.004;
\label{eq431}
\\
&&
\langle m_{Bl}  \rangle_{m_{Bl} > 50~{\rm GeV}}^{\rm NLO}
= 0.516 m_t - 0.96~{\rm GeV},\;\;\;\delta_{\rm rms} = 0.006,
\label{eq432}
\ea
where $\delta_{\rm rms}$ 
is the root mean square (rms) of the residuals of the linear 
fit. It is clear from the value of $\delta_{\rm rms}$ 
that the linear fit works 
very well. It is straightforward to translate the results of the linear
fit shown in Eqs.~(\ref{eq431},\ref{eq432}) to an {\it estimate}
of the error on the top quark mass. Indeed, suppose that a  typical
uncertainty of the measured value of $\langle m_{Bl}
\rangle $ is $0.4~{\rm GeV}$.
The slopes in Eqs.~(\ref{eq431},\ref{eq432}) then imply that
the corresponding error in the top quark mass $m_t$ is about
$0.8~{\rm GeV}$.  On the other hand, assuming perfect measurement
of $\langle m_{Bl} \rangle $, we find that theoretical 
uncertainties in $\langle m_{Bl} \rangle $ shown in Table~1 and the value 
of the slope of the linear fit  lead to a $1~{\rm GeV}$
uncertainty in the extracted value of the top quark mass.
The errors on the top quark mass that follow from the NLO QCD computation
are similar to differences between PYTHIA, HERWIG and NLO QCD.
To show this, we quote results of a
fit to $\langle m_{Bl} \rangle $  
obtained with PYTHIA and HERWIG in Ref. \cite{cmes}
\be
\langle m_{Bl} \rangle_{\rm Pythia} = 0.59~m_t - 24.11~{\rm GeV},
\;\;\;\;
\langle m_{Bl} \rangle_{\rm Herwig} = 0.61~m_t - 25.31~{\rm GeV}.
\label{eq351}
\ee
It is clear from the comparison of the fits Eqs.~(\ref{eq351},\ref{eq431})
that NLO QCD results
and parton shower results are close but not identical and these
differences are essential. Indeed,  we note that 
a slope difference between parton showers and NLO QCD is 
about $0.01$. Although such slope difference may look insignificant, 
it leads to  ${\cal O}(3~{\rm GeV})$ 
shift in the reconstructed value  of the top quark.  Hence,  parton showers are
insufficient for measurements of  the top quark mass with a precision
higher than a few GeV. On the contrary, it follows from 
Eqs.~(\ref{eq431},\ref{eq432}) 
that NLO QCD computations lead to results with small uncertainties
that can be estimated in a systematic way and, perhaps, be 
even further improved.

\subsection{Results: $m_{Bl}$ distribution
in $pp \to (t \to W^++b \to W^+ + B)  + ( \bar t \to W^- + \bar b)$
 }

In this Section, we consider production of $B$-mesons
through the fragmentation of $b$-quarks
in top decays but, in contrast to the previous Section,
we include the full production  process through next-to-leading order
in perturbative QCD. To claim that the NLO QCD computation can do a good
job in describing $\langle m_{Bl} \rangle $ in {\it reality},
it is very important to have
full production and decay chain included.

To this end, we consider top quark pair production in 
$pp$ collisions at $\sqrt{s} = 14~{\rm TeV}$ and focus on 
the lepton + jets decay channel.
We note that  NLO QCD corrections to the decay
$W \to q \bar q'$ need to be interfaced with $pp \to t \bar t$ production
process, to describe lepton + jets channel through NLO QCD.
We  require that there are at least 
four jets in the event. We include the $B$-meson
{\it and} the non-$B$-meson remnant of the fragmenting $b$-quark
into a list of proto-jets that are passed to the jet reconstruction algorithm.
We employ  $k_\perp$ jet algorithm with $R = 0.5$ and the
four-momentum recombination scheme. All reconstructed jets
and the positron from the $W^+$ decay are required to have  transverse momenta
in excess of $20~{\rm GeV}$\footnote{A standard argument \cite{cms,cmes}
that $\langle m_{Bl} \rangle$
involves a Lorentz invariant product of the two
four-vector and, therefore, does not depend on the production mechanism
is not applicable once cuts on the transverse momenta are applied. Such
cuts are only invariant under restricted class of Lorentz transformations -
boosts along the collision axis.}.  The
scalar sum of the transverse momenta of all jets
in the event should exceed  $100~{\rm GeV}$ \cite{cmsnote}. For the sake
of simplicity, we do not
impose any other kinematic constraints including cuts on the missing
energy and the lepton isolation cuts.  In addition,  we do not consider
combinatorial backgrounds, assuming that  the correct
pairing between a lepton and a $B$-meson can be established.
Finally, similar to what was done in Ref.~\cite{ms}, throughout this paper 
we consider 
intermediate top quarks to be on the mass-shell and we do not 
include the so-called non-factorizable corrections~\cite{my}. 
For observables  that we study in this paper, this is a good approximation 
since we, effectively, integrate over the invariant masses of each of the 
top quarks.

\begin{figure}[t!] 
\begin{center}
\scalebox{0.45}{\input{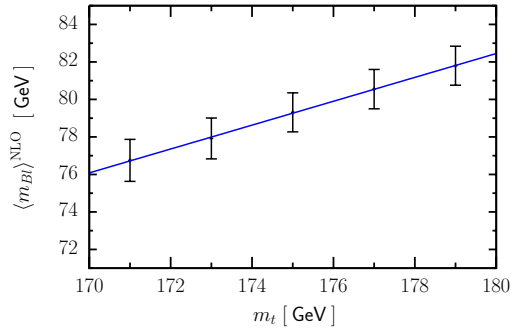}} 
\vspace{3mm}
\caption{Result of the linear fit to $\langle m_{Bl} \rangle^{\rm NLO}$
is  shown, {\it with all kinematic cuts on the final state particles 
applied}. See text for details. 
}
\label{fig25}
\end{center}

\end{figure}

In contrast to the previous Section, we do not change the parameters 
of the fragmentation functions, fixing them to their central 
values, see Eq.~(\ref{eq:ff_par}). We use CTEQ parton distribution 
functions \cite{Pumplin:2002vw,Nadolsky:2008zw} in the analysis.
For each input value of $m_t$ 
we compute $\langle m_{Bl} \rangle $ 
for three values of the renormalization and 
(pdf)-factorization scales $\mu_R = \mu_F = [m_t/4,m_t/2,m_t]$ and 
for three values of the factorization scale in the $b \to B$ fragmentation 
function $[m_t/4,m_t/2,m_t]$ and for two different types of fragmentation 
functions Eq.~(\ref{eq:ff_par}). As the result, for each value of $m_t$ we have 
eighteen values of $\langle m_{Bl} \rangle$ and $\sigma_{m_{Bl}}$. Although 
this is not an extensive scan of the parameter space, it gives a sense 
of theoretical  uncertainties in $\langle m_{Bl} \rangle $ provided 
that realistic production mechanism is employed\footnote{We have checked 
that if we only change parameters that are related  
to the decay process, we find  the 
${\cal O}(0.5~{\rm GeV})$ uncertainty in $\langle m_{Bl} \rangle$, similar 
to Tables 1,2.}. 
Computing 
the mean and the error, we arrive at 
the results shown in Table~3.  We see that the NLO QCD effects
in this case are quite small and {\it negative}, 
which is similar to the case of $m_{Bl} > 50~{\rm GeV}$ cut 
discussed earlier.  It is also clear that the uncertainty in $m_{Bl}$ 
decreases when NLO QCD corrections are included. 
By comparing results in Table~1 and in Table~3,
we see that effects of kinematic cuts on
$\langle m_{Bl} \rangle $ are  more important at leading order, where 
the average value of $m_{Bl}$ can shift by $3.5~{\rm GeV}$. On the other 
hand, at NLO, a typical shift is of the order of $0.8~{\rm GeV}$ 
and, therefore, is less dramatic.  Comparison of Tables~1 and~3 shows that 
the uncertainty of $\langle m_{Bl} \rangle $ at NLO nearly 
doubles if production 
mechanism is taken into account.  Performing the fit, we obtain (see Fig.\ref{fig25})
\be
\langle m_{Bl} \rangle^{\rm NLO}_{\rm prod} =
0.6365~m_t -32.12~{\rm GeV},\;\;\;\delta_{\rm rms} = 0.053.
\label{eq821}
\ee
Comparing this result with Eq.~(\ref{eq431}), we find a significant 
change in both the slope and the constant part. This demonstrates 
that $\langle m_{Bl} \rangle $ depends in a non-trivial way on the production 
mechanism, because of kinematic cuts applied to top quark decay products 
and additional jets in the production process.  It follows 
from Eq.~(\ref{eq821}) and the uncertainties of $\langle m_{Bl} \rangle $ 
shown in Table~3 that the theoretical error on the extracted value of $m_t$ 
is close to $1.5~{\rm GeV}$.


\begin{table}[t]
\label{Jtabb}
\vspace{0.1cm}
\begin{center}
\begin{tabular}{|c|c|c|c|c|}
\hline
$m_t$ &
$\langle m_{Bl}\rangle,\;{\rm LO}$ &
$\langle m_{Bl}\rangle,\;{\rm NLO}$ & $\sigma_{m_{Bl}},\; {\rm LO}$
& $\sigma_{m_{Bl}},\; {\rm NLO}$ \\ \hline \hline
$171$ &
$77.07 \pm 1.92$&
$76.75 \pm 1.12$ & 
$30.60 \pm 0.13$ & 
$28.41 \pm 0.36$ 
\\ \hline
$173$ &
$78.34 \pm 1.93$ &
$77.92 \pm 1.09$ & 
$31.01 \pm 0.14$ &
$28.72 \pm 0.31 $ 
\\ \hline
$175$ &
$79.63 \pm 1.98$ &
$79.31 \pm 1.04$ & 
$31.46 \pm 0.14$ &
$29.12 \pm 0.18$
\\ \hline
$177$ &
$80.91 \pm 2.03$  &
$80.55 \pm 1.05$  &
$31.83 \pm 0.15$  &
$29.48 \pm 0.13$
\\ \hline

$179$ &
$82.16 \pm 2.04$ &
$81.80 \pm 1.04$ &
$32.24 \pm 0.16$  &
$29.83 \pm 0.13$

 \\ \hline
\hline
\end{tabular}
\caption{The average values of the
invariant mass $\langle m_{Bl} \rangle$ and the dispersion 
in case where all the cuts on the final state particles are applied.
The top quark masses and all the results are in GeV.
See text for details.}
\vspace{-0.1cm}
\end{center}
\label{table:n3}
\end{table}

\section{Dilepton channel}
\label{dilepton}

In the  previous Section we saw that top quark decays to 
final states with identified hadrons
provide an interesting way to determine the top quark mass.
In this Section we study inclusive final states.
We focus on the  case where the top and the anti-top quarks
decay semileptonically, e.g. $t \to W^+ b \to l^+ \nu b$.  We
study the kinematic distribution of an invariant mass
of a  $b$-jet and a lepton, and the distributions of the sum of
energies of the two leptons and the two
$b$-jets. We employ
the NLO QCD corrections
to top quark pair production and decay, as computed in Ref.~\cite{ms}.
Throughout this Section, the center-of-mass energy of proton collisions 
is $14~{\rm TeV}$.

We begin by summarizing the kinematic cuts that are employed
to identify dilepton $t \bar t$ events \cite{Beneke:2000hk}.
Leptons are required to be central
$|\eta^{l}| < 2.5$ and have large transverse momentum
$p_\perp^{l} > 25~{\rm GeV}$. There should be  missing
energy in the event, $E_\perp^{\rm miss} > 40~{\rm GeV}$.
The jet transverse momentum cut is  $p_{\perp, j}> 25~{\rm GeV}$.
We employ the $k_\perp$ jet algorithm
with $R = 0.4$.

\subsection{Invariant mass of a lepton and a $b$-jet}

It is pointed out in Ref.~\cite{Beneke:2000hk} that
an average value of the invariant mass squared of a  $b$-jet and a
lepton $m_{lb}^2$ and an  average value of  the
the angle between the lepton and the  $b$-jet  in the $W$ boson rest frame,
can be used to construct an estimator of the top quark mass.
The estimator reads
\be
M_{\rm est}^2 = m_W^2 +  \frac{ 2\langle m_{lb}^2 \rangle}{1
- \langle \cos \theta_{lb} \rangle }.
\label{eq01}
\ee
To see that this is a good estimator, we note that 
for the top quark decay computed
at leading order in perturbative QCD
and without
any restrictions  on the final state $M_{\rm est}$ equals to $m_t$
\be
\langle m_{lb}^2 \rangle = \frac{m_t^2-m_W^2}{2}
\left ( 1 - \langle \cos \theta_{lb} \rangle \right ),\;\;\;
\langle \cos \theta_{lb} \rangle  = \frac{m_W^2}{m_t^2+ 2m_W^2}\;\;\;\;
\Rightarrow M_{\rm est}^2 = m_t^2.
\ee

In reality $M_{\rm est}$ is not equal to  $m_t$ for a variety
of reasons including i) kinematic
cuts required to identify the dilepton events; ii)  effects of higher
order QCD corrections; iii) impossibility to choose the ``correct''
pair of a lepton and a $b$-jet and iv) the experimental issues with
$b$-jet misidentification and the jet energy resolution. The computation
reported in Ref.~\cite{ms} allows us to calculate $M_{\rm est}^2$
within the framework of perturbative QCD, accounting for
the points i)-iii) {\it exactly}.

\begin{figure}[t]
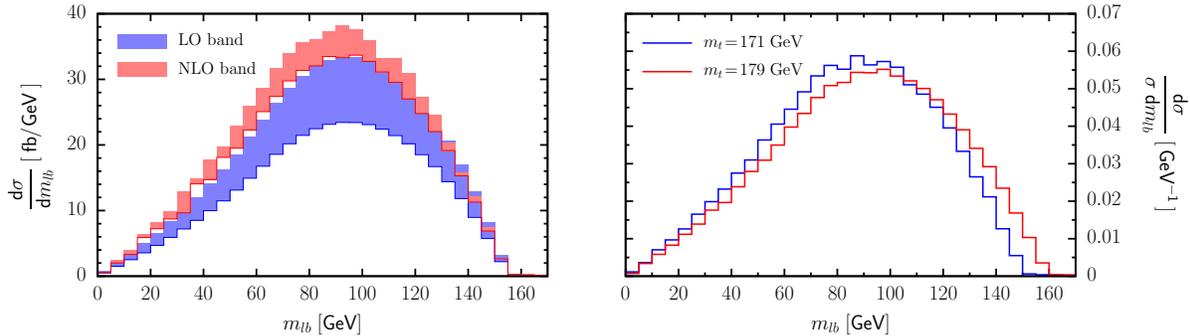
 
\begin{center}
  \scalebox{0.45}{\input{MLBScale.tex}} \hspace{5mm}
  \scalebox{0.45}{\input{MLB_mass.tex}}
\vspace{3mm}
\caption{The invariant mass distribution of the lepton
and the $b$-jet. Note that the lepton and the $b$-jet do not necessarily
come from the decay of the same top quark, see text.
The left  panel shows the scale uncertainty bands
for $\mu_R = \mu_F = [0.5m_t, 0.75m_t,m_t,1.25 m_t]$. The
right panel shows two NLO {\it normalized}
$m_{lb}$ distributions for $m_t = 171~{\rm GeV}$ and $m_t = 179~{\rm GeV}$.
}
\label{fig2}
\end{center}
\end{figure}

We point out that the computation of NLO QCD corrections to
$pp \to t \bar t$ process reported in \cite{ms} includes exact
spin correlations, one-loop effects in top quark decays and
allows arbitrary constraints to be imposed on  top quark decay products.
These features are crucial for reproducing experimental procedures.
Indeed, experimentally, it is not possible to determine  the
charge of the $b$-jet. Hence,  it is unclear
which of the two $b$-jets should be combined with the
chosen, definite-sign, lepton.
For the purpose of $m_{lb}$ reconstruction,
one pairs the lepton with the $b$-jet
that gives the smallest $m_{lb}$ value \cite{Beneke:2000hk}.
The parameter $\langle \cos \theta_{lb} \rangle $  in Eq.~(\ref{eq01})
is not measured and should be estimated theoretically. We have
also chosen to calculate $\langle \cos \theta_{lb} \rangle $ for the
$b$-jet that minimizes the invariant mass $m_{lb}$ since in this
case, there is a partial compensation of
incorrect assignments between the numerator and the denominator
in Eq.~(\ref{eq01}). As the result,
$M_{\rm est}$ becomes closer to the input value $m_t$ as compared
to the case when ``correct'' pairing of the  $b$-jet and the lepton
is chosen to calculate  $\langle \cos \theta_{lb} \rangle $
in Eq.~(\ref{eq01}).
It is argued in Ref.~\cite{Beneke:2000hk} that with $10~{\rm fb}^{-1}$
integrated luminosity, the statistical and systematic uncertainties
in the top quark mass
of about $1~{\rm GeV}$ each  can be achieved from
$\langle m_{lb}^2 \rangle $ measurement.

\begin{figure}[t!] 
\begin{center}
\scalebox{0.5}{\input{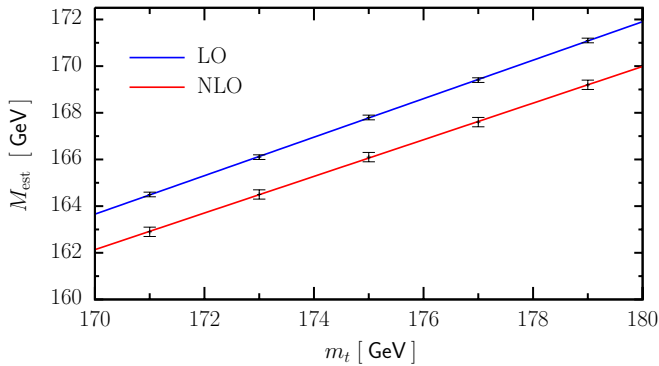}}
\vspace{3mm}
\caption{Results of a  linear fit to $M_{\rm est}$, Eq.~(\ref{eq01}), 
at leading and next-to-leading order in perturbative QCD.
}
\label{fig3}
\end{center}
\end{figure}

To assess how realistic those uncertainties are, we
consider five different values of the top quark
mass $m_t = [171,173,175,177,179]\;\;{\rm GeV}$. For each of these
$m_t$ values, we compute $M_{\rm est}$ for
four values of the renormalization and
the factorization scales
$\mu_R = \mu_F = [0.5m_t, 0.75m_t,m_t,1.25 m_t]$ and
for two sets of parton distribution functions CTEQ
\cite{Pumplin:2002vw,Nadolsky:2008zw}
and MRST \cite{Martin:2002aw}.
We use the mean value and the standard deviation
of these eight  values to compute central value of
$M_{\rm est}$ and its error.  Clearly, by no means
this is an exhaustive scan through the parameter
space\footnote{For example, one can and perhaps should
use different renormalization scales to compute
numerator and denominator in Eq.~(\ref{eq01}), to get a better
idea of the scale uncertainties in $M_{\rm est}$.}
 but it gives us an idea of the uncertainties on
the theoretical side.  Examples of $m_{lb}$ distributions and the results 
of the calculation are shown 
in Figs.~\ref{fig2},\ref{fig3}. The uncertainties on $M_{\rm est}$ do not depend 
on $m_t$ in significant way; they are $0.1~(0.2)~{\rm GeV}$ at leading 
and next-to-leading order, respectively.
Performing the linear fit, we find
\be
M_{\rm est}^{\rm LO}\;  =\; 0.8262 m_t + 23.22\;\mathrm{GeV},\;\;\;\;
M_{\rm est}^{\rm NLO}  = 0.7850 m_t + 28.70\;\mathrm{GeV}.
\label{eq010}
\ee

The quality of the 
linear fit is very good; for example,  the root mean square of the residuals
of the NLO fit is $\delta_{\rm rms} = 0.032$.
It is instructive that the analysis of this observable at leading
order shows stronger correlation between $m_t$ and $M_{\rm est}$ 
than at next-to-leading order. In addition, the theoretical 
uncertainty  in $M_{\rm est}$ increases when NLO QCD corrections are included.
The primary reason for the increased
uncertainty is stronger dependence of $M_{\rm est}$ on the
renormalization and factorization scales at NLO.  This feature can be understood
by considering the situation  where no phase-space cuts are applied and
where all the assignments of a lepton and a $b$-jet are done correctly.
In this case, as follows from the discussion at the beginning of this Section,
the estimator equals to the top quark mass regardless of the renormalization and
factorization scales and the chosen parton distribution functions.
 At next-to-leading order, this is not  true anymore because
of the gluon radiation in top decay
that is sensitive to the value of the strong coupling constant
and, hence, to the renormalization scale.
We note that we observe  a very weak dependence
of $M_{\rm est}$ on parton  distribution functions  which implies
that even with the phase-space cuts and incorrect pairing,
this variable is primarily sensitive to top quark decays rather
than to top quark  production mechanism.

Finally, we can use Eq.~(\ref{eq010}) and Fig.~\ref{fig3} to estimate
uncertainty in $m_t$ that can be achieved by measuring 
$M_{\rm est}$ with infinite precision.
Since, as follows from Fig.~\ref{fig3}, 
the uncertainty in $M_{\rm est}$ is $0.2~{\rm GeV}$ at NLO and given 
the slope of about $0.8$ in Eq.~(\ref{eq010}), we find 
the minimal uncertainty  in the extracted value of 
$m_t$ to be close to $0.25~{\rm GeV}$.
We note that  this result  does not include 
the $b$-quark fragmentation uncertainty 
and the jet scale uncertainty, estimated to be 
$0.7$ and $0.6~{\rm GeV}$, respectively,  in Ref.~\cite{Beneke:2000hk}. 

\subsection{Sum of energies of the two leptons from top quark decays}

Another observable that we consider is the sum of the energies
of the two leptons, $ E_{l_1}+E_{l_2} $, in the laboratory frame.  Lepton energies
in the laboratory frame can be easily measured and they are free
from jet  energy scale uncertainties that are important sources
of errors,  if the top quark mass is reconstructed from hadronic
final states. The important question is whether
or not the average value\footnote{One can ask the
same question about the shape of the distribution but such  discussion
is outside the scope of this paper.}
of the sum of  lepton energies is correlated
with the top quark mass at the parton level and how well such correlation
can be described by perturbative QCD.

\begin{figure}[t]
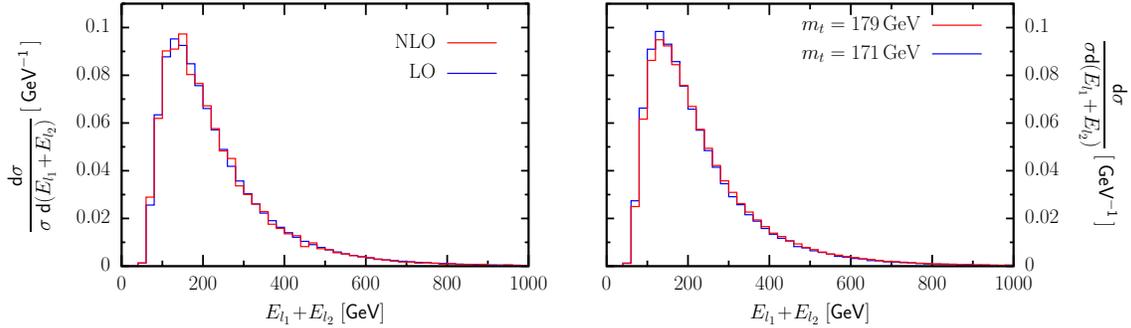
 
\begin{center}
  \scalebox{0.45}{\input{ELep.tex}}  \hspace{5mm}
  \scalebox{0.45}{\input{ELep_mass.tex}}
\vspace{3mm}
\caption{Left panel: normalized distribution of
the  sum of lepton energies at leading
and next-to-leading order calculated for
$m_t = 175~{\rm GeV}$. The renormalization and factorization
scales are set to $m_t$ 
and the MRST (left panel) and CTEQ (right panel)  
parton distribution functions set is used.
Note a shift in the position of the maximum of this distribution.
Right panel: normalized distributions of the sum of lepton energies
at next-to-leading order, for $m_t = 171~{\rm GeV}$ and
$m_t = 179~{\rm GeV}$.}
\label{fig4}
\end{center}
\end{figure}

\begin{figure}[t!] 
\begin{center}
  \scalebox{0.45}{\input{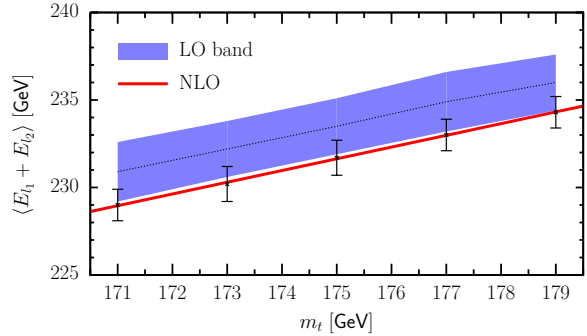}}
\vspace{3mm}
\caption{Results of a  linear fit to the sum of the average
energy of the two leptons
at leading and next-to-leading order.
}
\label{fig5}
\end{center}
\end{figure}

The corresponding distributions are shown in Fig.~\ref{fig4}. There we display
$E_{l_1}+E_{l_2}$ computed through leading and next-to-leading order
in perturbative QCD for $m_t = 175~{\rm GeV}$, as well the NLO QCD
distributions in $E_{l_1}+E_{l_2}$ for $m_t = 171$~GeV and $m_t =179~{\rm GeV}$.
To compute the mean value
of $\langle E_{l_1}+E_{l_2} \rangle $ we consider
the same range of the renormalization and factorization scales
and the two sets of  parton distribution functions, as in the previous
Section. The results of the calculation are shown in Fig.~\ref{fig5}. Performing a
linear fit,  we find
\ba
&& \langle E_{l_1}+E_{l_2} \rangle_{\rm LO}\;\;
= 0.645 m_t + 120.6\;\mathrm{GeV},\;\;\;\; \delta_{\rm rms} = 0.08;
\nonumber \\
&& \langle E_{l_1}+E_{l_2} \rangle_{\rm NLO}
= 0.670 m_t + 114.4\;\mathrm{GeV}, \;\;\;\; \delta_{\rm rms} = 0.07.
\label{eq1456}
\ea
The results of the linear fit are displayed in Fig.~\ref{fig5}. 
Theoretical errors on 
$\langle E_{l_1}+E_{l_2} \rangle$ are independent of the top mass;  
they are $1.7~{\rm GeV}$ at leading order 
and $1~{\rm GeV}$ at next-to-leading 
order.   Combining information about the slope in Eq.~(\ref{eq1456}) with 
the  theoretical uncertainty on 
$\langle E_{l_1}+E_{l_2} \rangle $, we conclude that the ultimate uncertainty 
in $m_t$ that can be achieved by studying this observable is close to 
$1.5~{\rm GeV}$.
It is interesting to point out that, in this case, both the scale
dependence of the NLO result and the difference between MRST and CTEQ
 parton distribution
functions are the two important sources of the uncertainty.

\subsection{Sum of jet energies}

Another observable that was discussed \cite{Beneke:2000hk}
in connection with  the top quark mass measurement is the sum of energies of
the two hardest  jets in the laboratory frame.
Similar to the  lepton energies just discussed,  the shape of the distribution
is an observable that is to be fitted; this is beyond the scope
of the present paper. Here, we limit ourselves to the discussion
of average values.  Instead of considering the
two hardest jets in the event, we found
it more useful to take the sum of energies of the two $b$-jets.

\begin{figure}[t] 
\begin{center}
  \scalebox{0.55}{\input{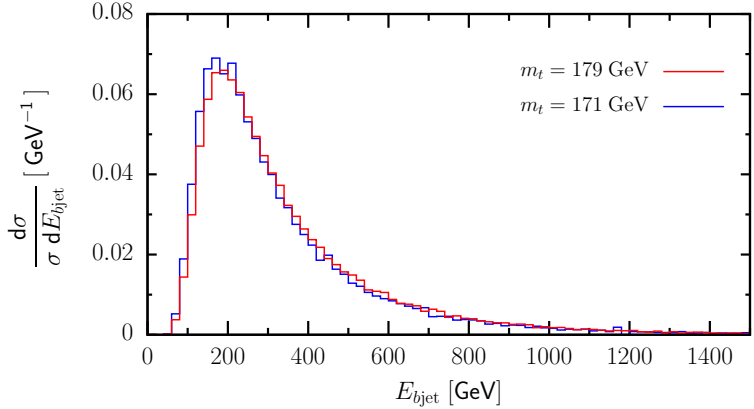}}
\vspace{3mm}
\caption{Normalized distribution of
the  sum of energies of two $b$-jets
$E_{\rm bjet} = E_{bj1}+E_{bj2}$
at next-to-leading order, calculated for
$m_t = 171~{\rm GeV}$ and
$m_t = 179~{\rm GeV}$. We set  renormalization and factorization
scales to $m_t$ and use CTEQ parton  distribution functions. }
\label{fig6}
\end{center}
\end{figure}

\begin{figure}[t!] 
\begin{center}
\scalebox{0.5}{\input{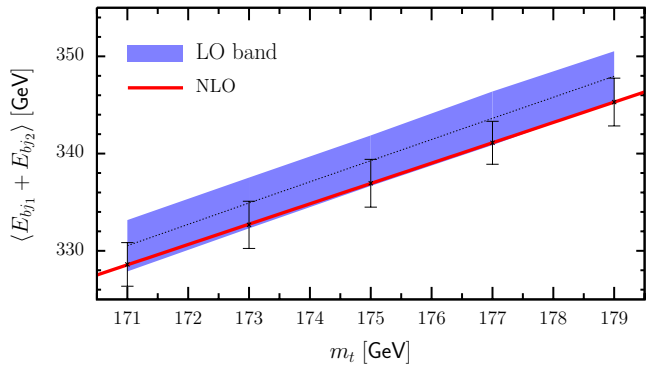}}
\vspace{3mm}
\caption{Results of a  linear fit to the sum of the average
energy of the two $b$-jets
at leading and next-to-leading order.
}
\label{fig7}
\end{center}
\end{figure}

We consider the distribution of
the sum of energies of the two $b$-jets. We expect that
this distribution is strongly correlated with the top quark mass, since
$b$-quarks originate directly from top decays. The results
of the calculation are shown in Figs.~\ref{fig6}.
Performing a linear
fit, we obtain the correlation between the average value of the
two $b$-jets and the top quark mass
\ba
&& \langle E_{bj1}+E_{bj2} \rangle_{\rm LO}
= 2.18  m_t -42.2\;\; {\rm GeV},\;\;\; \delta_{\rm rms} =0.02;
\nonumber \\
&& \langle E_{bj1}+E_{bj2} \rangle_{\rm NLO}
= 2.09 m_t -29.2\;\; {\rm GeV},\;\;\; \delta_{\rm rms} = 0.05.
\ea
The results of the linear fit together with theoretical uncertainties 
in $\langle E_{bj1}+E_{bj2} \rangle $ are shown in Fig.~\ref{fig7}.  
These uncertainties are $2.6~{\rm GeV}$ at LO  and $2.4~{\rm GeV}$ 
at NLO; they do not exhibit a strong dependence on the top quark mass.
Interestingly,  inclusion of NLO QCD corrections makes the
correlation between $\langle E_{bj1}+E_{bj2} \rangle_{\rm LO}$
and $m_t$ weaker.
However, the correlation is still quite   strong.  If we assume
that energies of $b$-jets  can be measured infinitely accurately,
the irreducible uncertainty on the determination of the 
top quark mass from $\langle E_{bj1}+E_{bj2} \rangle$
becomes only $1.2~{\rm GeV}$. 
Of course, the main issue here is to understand how well
$b$-jet energies  can actually be measured; 
this issue will be at the center of the experimental studies at the LHC.

\section{Conclusion}
\label{conc}

Determination of the top quark mass with high precision is
an important part  of the top quark physics program 
at the  LHC. It is expected, that a 
variety of methods will be employed  by 
ATLAS and CMS collaborations to measure the top quark mass. 
Some of those methods involve studies of the correlations
between the value of the top quark mass and the kinematics of the top quark
decay products. As the results of those studies, 
it is often claimed that the top quark mass can be determined 
with  ${\cal O}(1\%)$ uncertainty  at the LHC, but it is not clear 
whether or not these uncertainty   estimates  can be trusted.
Indeed, all such studies employ PYTHIA or HERWIG parton shower event
generators to describe top quark pair production and decay; however,  
no parton
shower is designed to handle this level of precision.
An interesting question therefore is to estimate, 
in a parton-shower-independent way,  the uncertainty on $m_t$ that
can be achieved in forthcoming LHC experiments. 

The goal of this paper is to address this question by computing
correlations between the top quark mass and the values of some 
 kinematic observables  through NLO QCD. Among other things,
we consider NLO QCD corrections to the invariant mass of a $B$-meson
and a lepton from top quark decays which is considered to be one of the
most accurate ways to determine the top quark mass.  Such computation
is rather unusual in the context of NLO QCD calculations  since it refers
to the final state with an identified hadron.

In general, we find that parton shower event generators do a good
job in estimating both the central value and the uncertainty in the top quark
mass that can be  achieved. However, as can be seen from the discussion
of the average value of the invariant mass of the $B$-meson and the lepton,
NLO QCD computations give both, a more accurate central values and an
estimate of the uncertainty that can be trusted.  Both of these features
are important if   we want to use the measured value of 
the top quark mass  with confidence, to constrain physics beyond the Standard Model 
through precision measurements.\\

{\bf Acknowledgments}
K.M. would like to acknowledge conversations with A.~Mitov that 
triggered this investigation and useful discussions with Z.~Kunszt. 
This research  is supported by the NSF under grant
PHY-0855365 and by the start up
funds provided by Johns Hopkins University.
Calculations reported in this paper were performed on the Homewood
High Performance Cluster of Johns Hopkins University.


\end{document}